\documentclass[10pt]{article}
\usepackage{bbm}
\usepackage[final]{graphics}
\usepackage{amsmath}
\usepackage{amsfonts,amsbsy}
\usepackage{amssymb}

\def\empile#1\over#2{\mathrel{\mathop{\kern 0pt#1}\limits_{#2}}}
\def\bs{\boldsymbol}
\def\wt#1{\widetilde{#1}}

\newcommand{\slv}{\raise.15ex\hbox{$/$}\kern-.53em\hbox{$v$}}
\newcommand{\slF}{\raise.15ex\hbox{$/$}\kern-.53em\hbox{$F$}}
\newcommand{\slL}{\raise.15ex\hbox{$/$}\kern-.53em\hbox{$L$}}
\newcommand{\slP}{\raise.15ex\hbox{$/$}\kern-.53em\hbox{$P$}}
\newcommand{\slp}{\raise.15ex\hbox{$/$}\kern-.53em\hbox{$p$}}
\newcommand{\slq}{\raise.15ex\hbox{$/$}\kern-.53em\hbox{$q$}}
\newcommand{\slR}{\raise.15ex\hbox{$/$}\kern-.53em\hbox{$R$}}
\newcommand{\slQ}{\raise.15ex\hbox{$/$}\kern-.53em\hbox{$Q$}}
\newcommand{\slK}{\raise.15ex\hbox{$/$}\kern-.53em\hbox{$K$}}
\newcommand{\slk}{\raise.15ex\hbox{$/$}\kern-.53em\hbox{$k$}}
\newcommand{\slD}{\raise.15ex\hbox{$/$}\kern-.53em\hbox{$D$}}
\newcommand{\slC}{\raise.15ex\hbox{$/$}\kern-.53em\hbox{$C$}}
\newcommand{\slA}{\raise.15ex\hbox{$/$}\kern-.53em\hbox{$A$}}
\newcommand{\slSigma}{\raise.15ex\hbox{$/$}\kern-.53em\hbox{$\Sigma$}}
\newcommand{\slpartial}{\raise.15ex\hbox{$/$}\kern-.53em\hbox{$\partial$}}
\newcommand{\slcalP}{\raise.15ex\hbox{$/$}\kern-.63em\hbox{$\cal P$}}

\def\p{{\boldsymbol p}}

\def\k{{\boldsymbol k}}

\def\x{{\boldsymbol x}}
\def\y{{\boldsymbol y}}

\catcode`\@=11

\newcount\@tempcntc
\def\@citex[#1]#2{\if@filesw\immediate\write\@auxout{\string\citation{#2}}\fi
  \@tempcnta\z@\@tempcntb\m@ne\def\@citea{}\@cite{%
        \@for\@citeb:=#2\do%
    {\@ifundefined{b@\@citeb}%
        {\@citeo\@tempcntb\m@ne\@citea%
                \def\@citea{,\penalty\@m\ }{\bf ?}\@warning%
                {Citation `\@citeb' on page \thepage \space undefined}}%
        {\setbox\z@\hbox{\global\@tempcntc0\csname b@\@citeb\endcsname\relax}
     \ifnum\@tempcntc=\z@ \@citeo\@tempcntb\m@ne%
       \@citea\def\@citea{,\penalty\@m}%
       \hbox{\csname b@\@citeb\endcsname}%
     \else%
      \advance\@tempcntb\@ne%
      \ifnum\@tempcntb=\@tempcntc%
      \else\advance\@tempcntb\m@ne\@citeo%
      \@tempcnta\@tempcntc\@tempcntb\@tempcntc\fi\fi}}\@citeo}{#1}}%

\def\@citeo{\ifnum\@tempcnta>\@tempcntb\else\@citea
  \def\@citea{,\penalty\@m}%
  \ifnum\@tempcnta=\@tempcntb\the\@tempcnta\else
   {\advance\@tempcnta\@ne\ifnum\@tempcnta=\@tempcntb \else
\def\@citea{--}\fi
    \advance\@tempcnta\m@ne\the\@tempcnta\@citea\the\@tempcntb}\fi\fi}

\catcode`\@=12

\begin{document}

\title{\bf How particles emerge from decaying classical
 fields in heavy ion collisions:\\
 \hskip -7mm
towards~a~kinetic~description~of~the~Glasma\!\!\!\!\!\!}
\author{Fran\c cois Gelis$^{(1)}$, Sangyong Jeon$^{(2)}$, Raju Venugopalan$^{(3)}$}
\maketitle
\begin{center}
\begin{enumerate}
\item Theory Division\\
  PH-TH, Case C01600, CERN,\\
  CH-1211 Geneva 23, Switzerland
\item Physics Department, McGill University,\\
  Montr\'eal, Qu\'ebec, H3A 2T8, Canada
\item Brookhaven National Laboratory,\\
  Physics Department, Nuclear Theory,\\
  Upton, NY-11973, USA
\end{enumerate}
\end{center}

\maketitle

\begin{abstract}
  
We develop the formalism discussed previously in hep-ph/0601209 and
hep-ph/0605246 to construct a kinetic theory that provides insight
into the earliest ``Glasma'' stage of a high energy heavy ion
collision. Particles produced from the decay of classical fields in
the Glasma obey a Boltzmann equation whose novel features include an
inhomogeneous source term and new contributions to the collision term.
We discuss the power counting associated with the different terms in
the Boltzmann equation and outline the transition from the field dominated
regime to the particle dominated regime in high energy heavy ion
collisions.

\end{abstract}
\vskip 5mm
\begin{flushright}
Preprint CERN-PH-TH/2007-106
\end{flushright}

\section{Introduction}
In two previous papers~\cite{GelisV2,GelisV3}, we introduced a
formalism to compute multi-particle production in field theories
coupled to strong time-dependent external sources. The QCD example of
such a field theory is the Color Glass Condensate
(CGC)~\cite{McLerV1,McLerV2,McLerV3,JalilKMW1,JalilKLW1,JalilKLW2,JalilKLW3,JalilKLW4,IancuLM1,IancuLM2,FerreILM1,McLer1,IancuLM3,IancuV1}. For
simplicity, we considered a $\phi^3$ theory; we believe however that
most of our results are of general validity and can be extended to
gauge theories~\cite{GelisLV1}.

In this paper, we will address a problem in multi-particle production
that was not considered in Refs.~\cite{GelisV2,GelisV3}. Specifically,
the approach developed there did not include scattering processes that
are important for the dynamics of the system at late times. These are
the so called {\sl secular terms} which are of higher order in the
coupling constant (loop corrections) and are accompanied by growing
powers of time~\cite{Golde1,VegaS1,BoyanV3}.  The secular
contributions must be resummed to obtain sensible results. In a
quantum field theory, this resummation is performed in principle by
solving the Dyson-Schwinger equations. In practice, the
Dyson-Schwinger equations are difficult to solve.  For a system of
fields coupled to an ensemble of particles, it is well known that the
Dyson-Schwinger equations can be approximated by a Boltzmann equation
for the distribution of particles. The goal of the present paper is to
extend the approach of Refs.~\cite{GelisV2,GelisV3} to derive a
kinetic equation that includes the late time contributions to
multi-particle production in field theories with strong external
sources. We have in mind the dynamics after a heavy ion collision,
where the classical field produced by the colliding nuclei expands
rapidly into the vacuum along the beam direction. Our approach may
also be of relevance to descriptions of the decay of the inflaton
field and thermalization in the preheating and reheating phases of the
early universe--a nice review with relevant references can be found in
Ref.~\cite{MichaT1}. In both cases, as the classical field evolves,
the occupation number decreases and it is more appropriate to describe
the higher momentum modes of the system in terms of particle degrees
of freedom.

The connections between the classical approximation in field theory
and kinetic equations in the framework of nuclear collisions were
previously discussed by Mueller and Son \cite{MuellS1}, and
subsequently by Jeon \cite{Jeon3}. They considered a system of fields
in the presence of an ensemble of particles described by a
distribution $f$. Performing a classical approximation in the path
integral describing the evolution of this system and a gradient
expansion in the obtained Dyson-Schwinger equations, these authors
obtained a kinetic equation for $f$.  An obvious question arises: with
what accuracy does this kinetic equation reproduce the Boltzmann
equation one would obtain without performing the classical
approximation?  The authors of Refs.~\cite{MuellS1,Jeon3} find that
the kinetic equation obtained from the classical path integral
reproduces correctly the collision term in the Boltzmann equation to
leading power of $f$ and (surprisingly) the first subleading term in
$f$ as well.
 
We shall adopt a more {\it ab initio} approach here by considering a
system that does not contain any particle degrees of freedom {\sl
initially}, but where the fields are coupled to a strong
time-dependent external source $j$. The external source is assumed to
be a stochastic variable that belongs to an ensemble of charges
specified by a distribution $W[j]$. This is the typical set up in the
description of heavy ion collisions in the Color Glass Condensate
framework where $W[j]$ represents the distribution of color
charges. Because of the expansion of the system, one may anticipate
that the system can be described by field theory methods at early
times and by kinetic theory and hydrodynamics at later times. The
matter in this regime in heavy ion collisions has interesting
properties; two noteworthy possibilities are dynamically generated
topological charge~\cite{KharzKV1,Shury1} and plasma instabilities
possibly leading to turbulent color fields~\cite{MrowcT1}. This
matter has been called an Glasma~\cite{LappiM1,GelisV4} and
understanding its dynamical evolution holds the key to a deeper
understanding of the strongly interacting Quark Gluon Plasma (sQGP)
that may be formed at later times~\cite{GyulaM2}. The 

We will address here general questions about the dynamical evolution
of such matter in the simplest possible context of a scalar ($\phi^3$)
field theory~\footnote{Even the ``simple" scalar theory is
non-trivial. It will indeed contain very general features of relevance
to the Glasma albeit the latter will have significant (and very
interesting) additional features that are absent in the scalar case.}:
\begin{itemize}
\item[{\bf i.}~] {\sl What is the kinetic equation one obtains in
  field theories coupled to strong external sources?}  Knowing the
  answer to this question is important for one to handle correctly the
  transition region between a field theory description and kinetic
  theory. Indeed, one expects from the work in
  Refs.~\cite{MuellS1,Jeon3} that there exists a window in time where
  both approaches correctly describe the dynamics~\footnote{This has
  to be the case if one wants the final result to be independent of
  the time at which one switches between the two descriptions.}. This
  suggests that the kinetic equation in the overlap regime must know
  about the coupling of sources to fields at earlier times. How is
  this manifest, how important is this effect and how does it go away
  ?
    
\item[{\bf ii.}~] {\sl What terms in the kinetic equation are
  important at different stages of the expansion? }  The previous
  question hints that we will obtain a kinetic equation that has
  additional terms absent in the conventional Boltzmann equation. We
  would like to understand how this generalized Boltzmann equation
  converges to the usual one at late times.
\end{itemize}

The paper is organized as follows. In section \ref{sec:nbar}, we shall
remind the reader of relevant formulae in the
derivation~\cite{GelisV2} of the average number $\big<n\big>$ of
produced particles. In section \ref{sec:dyson}, we shall write down
the Dyson-Schwinger equations for the two-point functions in theories
with time dependent strong sources. These provide the starting point
for a derivation in section \ref{sec:boltz} of the corresponding
kinetic equation for the Glasma. We observe that the coupling of the
field to an external source leads to an inhomogeneous term in this
kinetic equation. In section \ref{sec:discussion}, we discuss the
properties of the different terms appearing in the kinetic
equation. Albeit the collision term in the kinetic equation looks
identical to the collision term in the usual Boltzmann equation, it
contains novel contributions to the self energy that are of 0-loop and
1-loop order. We discuss the power counting for these different
contributions and assess their relative contribution at different
stages of the temporal evolution of the Glasma.  We conclude with a
brief summary and outlook emphasizing unresolved issues. 
An appendix addresses how the averaging over the
sources $j$ in our formalism can be re-expressed in terms of the usual
ensemble average implicit in the derivation of kinetic equations. 

\section{{\it Ab initio} computation of $\big<n\big>$}
\label{sec:nbar}
We consider the theory of a real scalar field $\phi$ with cubic
self-in\-te\-rac\-tions, coupled to an external time dependent source
$j(x)$. The Lagrangian of the model is
\begin{equation}
{\cal L}\equiv\frac{1}{2}\partial_\mu\phi\partial^\mu\phi 
-\frac{1}{2}m^2\phi^2-\frac{g}{3!}\phi^3 +j\phi\; .
\label{eq:lagrangian}
\end{equation}
In \cite{GelisV2}, we systematically calculated particle production
from these sources.  In the Color Glass Condensate
framework that this toy model mimics, the colliding projectiles are
represented by a statistical ensemble of currents $j$. Physical
quantities are obtained by averaging over all possible realizations of
the $j$'s. In this section, we shall discuss the calculation of
the average number of produced particles in a given configuration of
$j$'s.

A general formula for the average number $\big<n\big>$ of produced
particles is 
\begin{equation}
\big<n\big>=\int\frac{d^3\p}{(2\pi)^3 2E_p}\;
\big<0_{\rm in}\big|a^\dagger_{\rm out}(\p)a_{\rm out}(\p)\big|0_{\rm in}\big>
\; .
\end{equation}
The number of particles produced with a certain momentum $\p$ is
defined as the expectation value of the ``out'' number operator in the
initial state. This formula gives the number of particles at
asymptotic times, after the particles have decoupled~\footnote{The
``number of particles'' at some intermediate time, while the fields
are still interacting, is not a well defined concept.}.

A simple reduction formula gives~\cite{ItzykZ1}
\begin{eqnarray}
&&
\big<0_{\rm in}\big|a^\dagger_{\rm out}(\p)a_{\rm out}(\p)\big|0_{\rm in}\big>
=
\frac{1}{Z}\int d^4x\, d^4y\; e^{-ip\cdot x} e^{ip\cdot y}
\nonumber\\
&&\qquad\qquad\qquad\qquad\times
(\square_x+m^2)(\square_y+m^2)\;
\big<0_{\rm in}\big|\phi(x)\phi(y)\big|0_{\rm in}\big>\; ,
\label{eq:reduc-nbar}
\end{eqnarray}
where $Z$ is the wave function remormalization factor.  The
expectation value in the right hand side of this equation has two
important features~: (i) the vacuum state is the ``in'' vacuum state
on both sides and, (ii) the two fields inside the correlator are not
time-ordered. The Schwinger-Keldysh formalism \cite{Schwi1,Keldy1} provides 
techniques for computing these types of correlators. 

The operators $\square+m^2$ amputate the external legs of the
two-point function $G_{-+}(x,y)\equiv\big<0_{\rm
in}\big|\phi(x)\phi(y)\big|0_{\rm in}\big>$. Defining
\begin{equation}
\widetilde{G}_{-+}(x,y)\equiv
\frac{(\square_x+m^2)(\square_y+m^2)}{Z}\;G_{-+}(x,y)\; ,
\end{equation}
we can write the average multiplicity as
\begin{equation}
\big<n\big>=\int\frac{d^3\p}{(2\pi)^3 2E_p}\;
\int d^4x \,d^4y\; e^{-ip\cdot x} e^{ip\cdot y}\;
\widetilde{G}_{-+}(x,y)\; .
\label{eq:nbar-wigner}
\end{equation}
Introducing the variables
\begin{equation}
X\equiv\frac{x+y}{2}\; ,\quad
r\equiv x-y\; ,
\end{equation}
we can rewrite this formula as
\begin{equation}
E_\p\frac{d\big<n\big>}{d^3\p}=\frac{1}{16\pi^3} \int d^4X \;
\widetilde{G}_{-+}(X,p)\; ,
\label{eq:nbar1}
\end{equation}
where 
\begin{equation}
\widetilde{G}_{-+}(X,p)\equiv
\int d^4r \;
e^{-ip\cdot r}\;
\widetilde{G}_{-+}\left(X+\frac{r}{2},X-\frac{r}{2}\right)
\label{eq:wigner-def}
\end{equation}
is the Wigner transform of $\widetilde{G}_{-+}(x,y)$.

In the Schwinger--Keldysh formalism, the propagators ${\bs G}_{\epsilon\epsilon^\prime}(x,y)$, ($\epsilon,\epsilon^\prime= +,-$) can be 
expressed as 
\begin{equation}
{\bs G}_{\epsilon\epsilon^\prime}(x,y)=
\frac{\delta}{i\delta j_\epsilon(x)}\,
\frac{\delta}{i\delta j_{\epsilon^\prime}(y)}\;
\left.
e^{i{\cal V}_{_{SK}}[j_+,j_-]}
\right|_{j_+=j_-=j}\; ,
\label{eq:G-full}
\end{equation}
where $i{\cal V}_{_{SK}}[j_+,j_-]$ is the sum of all {\bf connected}
vacuum-vacuum diagrams. When $j_+=j_-=j$, $i{\cal
V}_{_{SK}}[j,j]=0$ and the sum of all vacuum-vacuum diagrams is unity.

Working out the functional derivatives,
\begin{equation}
{\bs G}_{\epsilon\epsilon^\prime}(x,y)
=\left[
\frac{\delta i{\cal V}_{_{SK}}}{i\delta j_\epsilon(x)}
\frac{\delta i{\cal V}_{_{SK}}}{i\delta j_{\epsilon^\prime}(y)}
+
\frac{\delta^2 i{\cal V}_{_{SK}}}{i\delta j_\epsilon(x)i\delta j_{\epsilon^\prime}(y)}
\right]_{j_+=j_-=j}\; .
\label{eq:G0}
\end{equation}
\begin{figure}[htbp]
\begin{center}
\resizebox*{5cm}{!}{\includegraphics{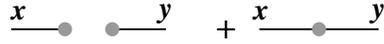}}
\end{center}
\caption{\label{fig:G} Diagrammatic representation of the disconnected
(left) and connected (right) terms in eq.~(\ref{eq:G0}). The gray
blobs denote the remnants of Green's functions after the free
propagators at the endpoints are amputated.}
\end{figure}
As $i{\cal V}_{_{SK}}$ is the sum of connected vacuum-vacuum
diagrams, any of its derivatives with respect to $j_\pm$ is a
connected Green's function. Therefore, 
${\bs G}_{\epsilon\epsilon^\prime}$ can be decomposed as 
\begin{equation}
{\bs G}_{\epsilon\epsilon^\prime}(x,y)
\equiv
{\bs G}^{\rm c}_{\epsilon\epsilon^\prime}(x,y)
+
{\bs G}^{\rm nc}_{\epsilon\epsilon^\prime}(x,y)\; .
\label{eq:G}
\end{equation}
These are, respectively, the connected part
\begin{equation}
{\bs G}_{\epsilon\epsilon^\prime}^{\rm c}(x,y)\equiv
\left. 
\frac{\delta^2 i{\cal V}_{_{SK}}}{i\delta j_\epsilon(x)i\delta j_{\epsilon^\prime}(y)}\right|_{j_+=j_-=j}\; ,
\end{equation}
and a disconnected part corresponding to the product of the
expectation values of the field at the points $x$ and $y$~:
\begin{equation}
{\bs G}^{\rm nc}_{\epsilon\epsilon^\prime}(x,y)=
\big<\phi(x)\big>\big<\phi(y)\big>\quad \mbox{with}\quad
\big<\phi(x)\big>=\left. 
\frac{\delta i{\cal V}_{_{SK}}}{i\delta j_\pm(x)}
\right|_{j_+=j_-=j}\; .
\end{equation}
When $j_+=j_-=j$, the expectation value of the field is the
same on the upper and lower branches of the contour:
$\big<\phi_+(x)\big>=\big<\phi_-(x)\big>$. This explains why we
 omitted the $+/-$ index in the expectation value of the field.
 
\begin{figure}[htbp]
\begin{center}
\resizebox*{!}{5cm}{\rotatebox{90}{\includegraphics{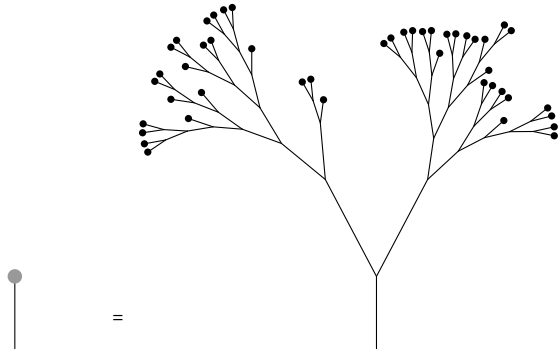}}}
\end{center}
\caption{\label{fig:tree}Example of a tree diagram contributing to the
field expectation value. The black dots terminating branches of the
tree represent insertions of the source $j$ in the diagram on the
right. The sum of these tree diagrams is represented (left) by a line
attached to a gray blob.}
\end{figure}

A typical tree-level contribution to $\big<\phi(x)\big>$ is shown
in figure \ref{fig:tree}.
Note also that $\big<\phi(x)\big>$ vanishes if the external source
$j(x)$ is zero\footnote{We assume that the self-interactions of the
fields are such that there is no spontaneous breakdown of symmetry
when $j=0$.}. At tree level, because
$j_+=j_-=j$, the sum over the $+/-$ indices in the Schwinger-Keldysh
formalism at all the internal vertices of the tree (including the
sources) can be performed by using the identities
\begin{eqnarray}
{\bs G}^0_{++}-{\bs G}^0_{+-}={\bs G}^0_{_R}\quad,\quad
{\bs G}^0_{-+}-{\bs G}^0_{--}={\bs G}^0_{_R}\; ,
\label{eq:G-ret}
\end{eqnarray}
where ${\bs G}^0_{_R}$ is the free retarded propagator\footnote{In
momentum space, this propagator reads ${\bs
G}^0_{_R}(p)=i/(p^2-m^2+ip^0\epsilon)$.}.  When this sum is performed,
all propagators in the tree diagram can be simply replaced by retarded
propagators. This is equivalent to the statement that
$\big<\phi(x)\big>$ is the retarded solution of the classical equation
of motion,
\begin{equation}
(\square+m^2)\phi(x) +\frac{g}{2}\phi^2(x) = j(x)\; ,
\label{eq:classEOM}
\end{equation}
with a vanishing boundary condition at $x_0=-\infty$.

Eq.~(\ref{eq:nbar1}) is the complete answer to the problem of particle
production in the effective theory described by the Lagrangian of
eq.~(\ref{eq:lagrangian}). If one were able to compute
$\widetilde{G}(x,y)$ to all orders, this formula would contain
everything one needs. There would be no need for tools such as kinetic
theory.

However, evaluating eq.~(\ref{eq:nbar1}) to all orders is an
unrealistic goal. What has been implemented thus far is
the evaluation of eq.~(\ref{eq:nbar1}) at leading order (tree level)
to calculate the gluon yield in high-energy nucleus-nucleus collisions
\cite{KrasnV4,KrasnV1,KrasnV2,KrasnNV1,KrasnNV2,Lappi1}.  In
\cite{GelisV2}, an algorithm was sketched to compute $\big<n\big>$ at
next-to-leading order (one loop) in terms of the retarded classical
field and of retarded fluctuations propagating in the classical field
background.

In practice, one has to truncate the loop expansion. As we will
discuss in the next section, the correct way to perform practical
calculations is within the framework of the Dyson--Schwinger
equations.

\section{Dyson-Schwinger equations}
\label{sec:dyson}
The main problem with the loop expansion described in the previous
section is that, in general, truncations in $\widetilde{G}(x,y)$ will
lead to an incorrect large time limit of the number of produced
particles. This can be traced to secular terms containing powers of
the time that invalidate the perturbative series in the large time
limit.  This can be cured by appropriate resummation; the well known
way to do this is to solve Dyson-Schwinger
equations~\cite{Golde1,VegaS1,BoyanV3}. In this section, we
shall discuss the Dyson-Schwinger equations obeyed by the two-point
functions $G_{\epsilon\epsilon^\prime}(x,y)$ of the Schwinger-Keldysh
formalism. We will see that the presence of a disconnected
contribution to these 2-point functions leads to interesting features
in the corresponding Dyson-Schwinger equations.

\subsection{Dyson-Schwinger equation for the connected part}
It is straightforward to write a Dyson-Schwinger equation for the
connected part of the 2-point function, ${\bs G}^{\rm
c}_{\epsilon\epsilon^\prime}$, that resums self-energy corrections~:
\begin{equation}
{\bs G}^{\rm c}(x,y)={\bs G}^0(x,y)+\int_{\cal C}d^4u\, d^4v\;
{\bs G}^0(x,u)\Big[-i{\bs \Sigma}(u,v)\Big]{\bs G}^{\rm c}(v,y)\; ,
\label{eq:DSc0}
\end{equation}
where $-i{\bs \Sigma}$ is a 1-particle irreducible
connected\footnote{It is connected in order to have a connected
2-point function after the resummation and it needs to be 1PI to
prevent double counting.} self-energy, {\sl evaluated in the presence
of external sources}. We shall not write here explicitly the $\pm$
indices carried by the various objects. Instead, we write the time
integrations as integrals over the complete Schwinger-Keldysh contour
${\cal C}$.
\begin{figure}[htbp]
\begin{center}
\resizebox*{6.8cm}{!}{\includegraphics{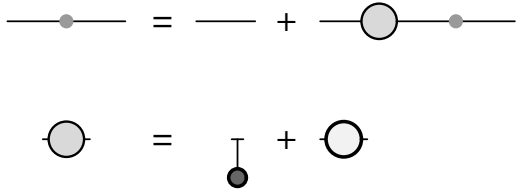}}
\end{center}
\caption{\label{fig:DSc0} Top: Diagrammatic representation of the
Dyson-Schwinger equation of eq.~(\ref{eq:DSc0}). The large gray blob
denotes the 1-particle irreducible 2-point function
${\bs\Sigma}$. Bottom: decomposition of the 1PI self-energy
${\bs\Sigma}$ into a local part $g{\bs\Phi}$ and a non-local part
${\bs\Pi}$ (denoted by a large light-gray blob), following
eq.~(\ref{eq:sigma-split}).}
\end{figure}

It is convenient to extract from this self-energy a local piece, by
writing
\begin{equation}
{\bs \Sigma}(u,v)
\equiv
g{\bs \Phi}(u)\delta(u-v)+{\bs \Pi}(u,v)\; .
\label{eq:sigma-split}
\end{equation}
Except for the background field, which is a genuine local contribution
to the self-energy, there is a certain arbitrariness in this
separation because it depends on the momentum scale at which we
resolve the system. A contribution to the self-energy that does not
change significantly over space-time scales on the order of the
Compton wavelength $p^{-1}$ can be treated as a mean field at that
scale. Therefore, the mean field term ${\bs \Phi}(u)$ will contain the
classical field, and possibly changes in the dispersion relation due
to medium effects~\footnote{To allow for this possibility, we denote
the mean field piece by a symbol distinct from the one used for the
classical field.}.

The Dyson-Schwinger equation then becomes
\begin{eqnarray}
{\bs G}^{\rm c}(x,y)&=&{\bs G}^0(x,y)
-ig
\int_{\cal C}d^4u \;
{\bs G}^0(x,u){\bs \Phi}(u){\bs G}^{\rm c}(u,y)
\nonumber\\
&&\qquad\qquad
+
\int_{\cal C}d^4u d^4v\;
{\bs G}^0(x,u)\Big[-i{\bs \Pi}(u,v)\Big]{\bs G}^{\rm c}(v,y)\; .
\end{eqnarray}
Using 
\begin{equation}
\big[\square_x+m^2\big]{\bs G}^0(x,y)=-i\delta_{_{\cal C}}(x-y)\; ,
\end{equation}
where $\delta_{_{\cal C}}$ denotes the delta function on the closed
time path\footnote{$\delta_{_{\cal C}}(x-y)=0$ unless $x^0$ and $y^0$
are equal and lie on the same branch of the time path.}, we can
rewrite this equation as
\begin{eqnarray}
\big[
\square_x+m^2+g{\bs \Phi}(x)
\big]{\bs G}^{\rm c}(x,y)=-i\delta_{_{\cal C}}(x-y)
-
\int_{\cal C}d^4u\; {\bs \Pi}(x,u){\bs G}^{\rm c}(u,y)\; .
\label{eq:C1}
\end{eqnarray}

\subsection{Dyson-Schwinger equation for the disconnected part}
We also need a Dyson-Schwinger equation for the disconnected part of
the Green's function,
\begin{equation}
{\bs G}^{\rm nc}(x,y)
=
\big<\phi(x)\big>\big<\phi(y)\big>\; .
\end{equation}
Because the expectation value $\big<\phi\big>$ is a connected 1-point
function, it is natural to factor the connected propagator out of
it, by writing
\begin{equation}
\big<\phi(x)\big>\equiv
\int_{\cal C}d^4u\;
{\bs G}^{\rm c}(x,u){\bs S}(u)\; ,
\label{eq:S-def}
\end{equation}
where ${\bs S}(u)$ is an ``effective source" term\footnote{In the
classical limit, one has ${\bs S}(x)=j(x)+\frac{g}{2}\phi^2(x)$ (see
section \ref{sec:source_term}).}. By construction, one obtains
\begin{equation}
\big[
\square_x+m^2+g{\bs \Phi}(x)
\big]
\big<\phi(x)\big>
=
-i{\bs S}(x)
-
\int_{\cal C}d^4u\; {\bs \Pi}(x,u)\big<\phi(u)\big>\; .
\end{equation}

Multiplying both sides by $\big<\phi(y)\big>$, one obtains
\begin{equation}
\big[
\square_x+m^2+g{\bs \Phi}(x)
\big]{\bs G}^{\rm nc}(x,y)
=
-i{\bs S}(x)\big<\phi(y)\big>
-
\int_{\cal C}d^4u\; {\bs \Pi}(x,u){\bs G}^{\rm nc}(u,y)\; .
\end{equation}
Defining
\begin{equation}
-i{\bs \Pi}^{_S}(x,y)\equiv {\bs S}(x){\bs S}(y)\; ,
\label{eq:pi-source}
\end{equation}
we can rewrite this equation as
\begin{equation}
\big[
\square_x+m^2+g{\bs \Phi}(x)
\big]{\bs G}^{\rm nc}(x,y)
=
-
\int_{\cal C}d^4u\;
\Big[
 {\bs \Pi}^{_S}(x,u){\bs G}^{\rm c}(u,y)
+
{\bs \Pi}(x,u){\bs G}^{\rm nc}(u,y)
\Big]
\; .
\label{eq:NC1}
\end{equation}
Adding eqs.~(\ref{eq:C1}) and (\ref{eq:NC1}), we obtain the
Dyson-Schwinger equation for the complete two-point function:
\begin{eqnarray}
&&
\big[
\square_x+m^2+g{\bs \Phi}(x)
\big]{\bs G}(x,y)
=
-i\delta_{_{\cal C}}(x-y)
\nonumber\\
&&\qquad\qquad\qquad\qquad
-
\int_{\cal C}d^4u\;
\Big[
 {\bs \Pi}^{_S}(x,u){\bs G}^{\rm c}(u,y)
+
{\bs \Pi}(x,u){\bs G}(u,y)
\Big]
\; .
\label{eq:CNC1}
\end{eqnarray}
The only formal difference between this Dyson-Schwinger
equation and the equation one obtains in the absence of the
source $j$ is the term proportional to ${\bs \Pi}^{_S}$ in the right
hand side.

In principle, the resummations performed by solving eqs.~(\ref{eq:C1})
and (\ref{eq:NC1}) (or, equivalently, eqs.~(\ref{eq:C1}) and
(\ref{eq:CNC1})) would completely cure the problem of secular
terms. Such an approach has been pursued numerically in \cite{Berge2},
but has not been attempted yet in the context of heavy ion collisions
in the CGC framework.

\section{Kinetic equation}
\label{sec:boltz}
The Dyson-Schwinger equations we wrote down in the previous section
contain all the necessary physics but their solution is likely too
difficult; they therefore by themselves do not provide any practical
insight into the dynamics of high energy heavy ion collisions.  One
can simplify the problem a step further by transforming the
Dyson-Schwinger equations for the 2-point functions into kinetic
equations. However, as we shall discuss shortly, doing so requires
that certain assumptions be satisfied.

\subsection{Fields and particles}
As is well known, the Boltzmann kinetic equation describes the
space--time evolution of particle phase space densities.  Therefore,
to achieve a kinetic description, the formalism considered thus far
should be extended to incorporate an ensemble of particles.  This is
simply done by modifying the free propagators to add a term that
depends on the distribution of particles $f(\p)$. In momentum space,
the modified propagators are\footnote{The propagators of the
Schwinger-Keldysh formalism appropriate for calculating
eq.~(\ref{eq:reduc-nbar}) are the same with $f(\p)=0$.}
\begin{eqnarray}
&&
{\bs G}^0_{++}(p)\equiv \frac{i}{p^2-m^2+i\epsilon}
+2\pi f(\p)\delta(p^2-m^2)
\; ,
\nonumber\\
&&
{\bs G}^0_{--}(p)\equiv 
\frac{-i}{p^2-m^2-i\epsilon}
+2\pi f(\p)\delta(p^2-m^2)
\; ,
\nonumber\\
&&
{\bs G}^0_{-+}(p)\equiv 
2\pi (\theta(p^0)+f(\p))\delta(p^2-m^2)
\; ,
\nonumber\\
&&
{\bs G}^0_{+-}(p)\equiv 
2\pi (\theta(-p^0)+f(\p))\delta(p^2-m^2)
\; .
\label{eq:SK-f}
\end{eqnarray}

These modified rules for the Schwinger--Keldysh propagators can be
derived~\cite{Bella1} when the {\sl initial} density matrix that
describes the ensemble has the form
\begin{equation}
\rho\equiv\exp\Big[
-\int\frac{d^3\p}{(2\pi)^3 2E_\p}\;\beta_\p E_\p\;
a_{\rm in}^\dagger(\p)a_{\rm in}(\p)
\Big]\; ,
\end{equation}
where $\beta_\p$ is a momentum dependent quantity. (Note: $\beta_\p$ should not be 
confused with the inverse temperature.) Such a form for the density matrix is required if
correlators computed with this density matrix are to satisfy Wick's
theorem. From this form of the density matrix, one obtains the
Schwinger-Keldysh rules of eqs.~(\ref{eq:SK-f}), with
\begin{equation}
f(\p)=\frac{1}{e^{\beta_\p E_\p}-1}\; .
\end{equation}
The function $f(\p)$ in the propagators only represents the {\sl
initial} distribution of particles in the system. Thus the field
theory defined by the Lagrangian of eq.~(\ref{eq:lagrangian}) and the
propagators of eqs.~(\ref{eq:SK-f}) describes a system of fields
coupled to an external source $j$ and to an ensemble of particles with
an initial distribution $f(\p)$. The Feynman rules then enable one to
calculate the properties of this system at a later time.

However, eqs.~(\ref{eq:SK-f}) do not lead to a well behaved
perturbative expansion, except when the function $f(\p)$ is the
equilibrium Bose-Einstein distribution in our model of bosonic fields.
In general, when $f(\p)$ is not a Bose-Einstein distribution, the
perturbative expansion based on eqs.~(\ref{eq:SK-f}) is plagued by the
previously mentioned pathological secular terms which need to be
resummed.  The time-scale at which resummation becomes necessary is
related to the transport mean free path in the system, namely, the
time between two large angle scatterings undergone by a particle. This
resummation makes the distribution $f(\p)$ time--dependent
reflecting the changes induced by collisions on the particle phase space
distribution. Under certain approximations to be discussed later, this
temporal evolution is governed by a Boltzmann equation.

The problem formulated in section \ref{sec:nbar} concerned a system
that has no ensemble of particles at the initial time ($f(\p)=0$ in
eqs.~(\ref{eq:SK-f})). At first sight, as $f=0$ is a particular case
of the Bose--Einstein distribution (with a vanishing temperature),
secular divergences may appear to be absent.  However, this conclusion
is incorrect because of the presence of external sources which drive
the system out of equilibrium. Thus it is also necessary to resum
secular terms in this case, leading to changes in $f(\p)$. The
generalized propagators in eq.~(\ref{eq:SK-f}) constitute the natural
framework to achieve this. Because the external source is both time
and space dependent, one has more generally
\begin{equation}
f(\p)\quad\to\quad f(X,\p)
\end{equation}
in eqs.~(\ref{eq:SK-f}).

An important point must be made here about the tree level expectation
values $\big<\phi_\pm(x)\big>$ in this f-dependent extension of our
formalism. A crucial property of the propagators in
eqs.~(\ref{eq:SK-f}) is that they still obey
eqs.~(\ref{eq:G-ret}). The retarded propagator is therefore {\sl
$f$-independent.}  Therefore, as long as loop corrections are not
included, the field expectation value does not depend on $f$ and is
identical to the result obtained from the retarded solution of the
classical equations of motion. Hence, the contribution from the
disconnected part of the 2-point function lead to an inhomogeneous
($f$--independent) term in the Boltzmann equation.

\subsection{Gradient expansion}
The extension (\ref{eq:SK-f}) of the propagators leads to
Dyson-Schwinger equations that are formally identical to
eqs.~(\ref{eq:C1}) and (\ref{eq:NC1}) -- with all the building blocks
now constructed with $f$-dependent propagators. The first step in
obtaining the Boltzmann equation is to rewrite all the distributions
in terms of their Wigner transforms. For a two-point function
$F(x,y)$, its Wigner transform ${\tilde F}(X,p)$ is defined to be
\begin{equation}
{\tilde F}(X,p)\equiv \int d^4s\; e^{-ip\cdot s}\;
F\left(X+\frac{s}{2},X-\frac{s}{2}\right)\; .
\end{equation}

The next step is to perform a gradient expansion where only long
wavelength, low momentum modes are retained. In particular, all terms
of order two or higher in $\partial_{_X}$ are neglected.  As our goal
is to construct a kinetic theory for the Glasma, we will discuss the
validity of this gradient expansion in the context of heavy ion
collisions in the CGC framework.  In this
framework~\cite{McLer1,IancuLM3,IancuV1}, the color sources
$\rho^a(\x_\perp)$ generating the color currents~\footnote{These color sources are the QCD analogs of the sources $j$ in 
our toy scalar theory.} are stochastic
variables that vary from event to event with a distribution
$W[\rho]$. When calculating a given physical quantity, one first
computes it for an arbitrary $\rho$ and then averages over all
possible $\rho$'s in the ensemble generated with the weight
$W[\rho]$. For example, in the McLerran-Venugopalan model
\cite{McLerV1,McLerV2,McLerV3}, the distribution $W[\rho]$ is a
Gaussian with
\begin{equation}
W[\rho]=\exp\Big(-\int d^2\x_\perp d^2\y_\perp\;
\frac{\rho(\x_\perp)\rho(\y_\perp)}{2\,\mu^2(\x_\perp,\y_\perp)}\Big)\; ,
\label{eq:MV}
\end{equation}
where 
\begin{equation}
\mu^2(\x_\perp,\y_\perp)\equiv 
\big<\rho(\x_\perp)\rho(\y_\perp)\big>
=\mu_{_A}^2(\x_\perp)\delta(\x_\perp-\y_\perp)\; .
\label{eq:2p-corr}
\end{equation}
Here $\mu_{_A}^2(\x_\perp)$ represents the density of color charges at a 
spatial position $\x_\perp$ in the nucleus.  The typical momentum scale of the sources--the saturation momentum squared
$Q_s^2$ at $\x_\perp$ is simply related to $\mu_{_A}^2$.

\begin{figure}[htbp]
\begin{center}
\resizebox*{5.6cm}{!}{\includegraphics{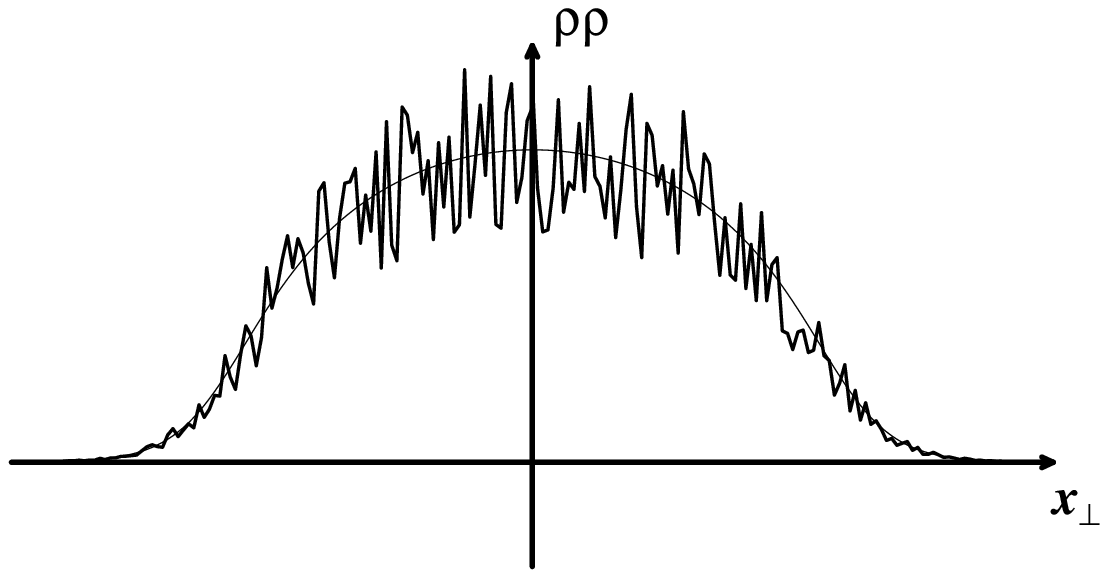}}
\hfill
\resizebox*{5.6cm}{!}{\includegraphics{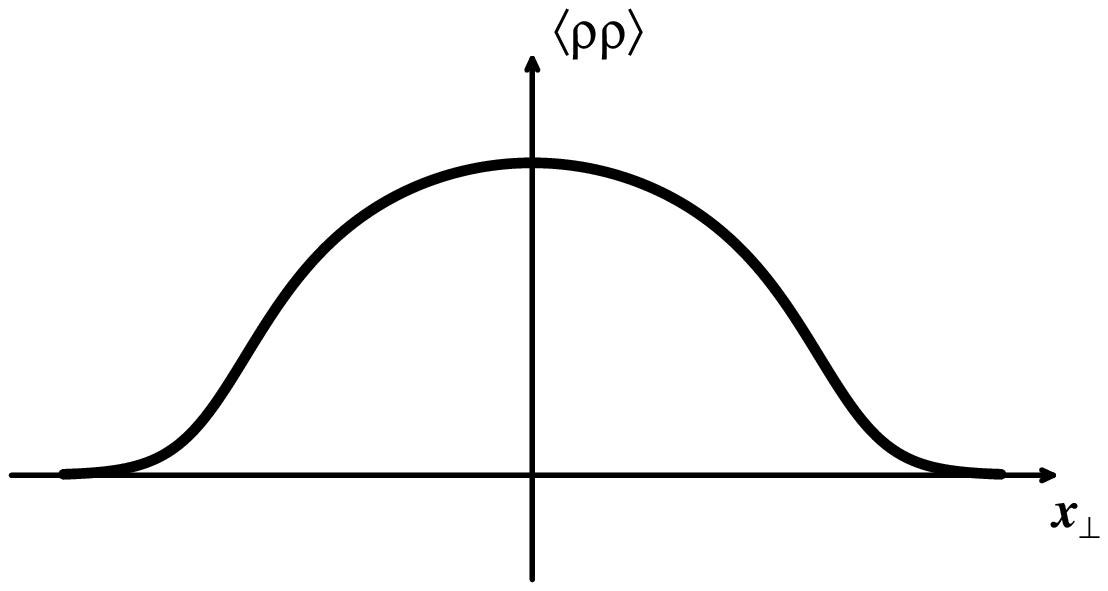}}
\end{center}
\caption{\label{fig:fluct} Left: $\rho^2$ distribution for one
  configuration in the ensemble represented by the distribution
  $W[\rho]$.  Right: ensemble average of $\left<\rho^2\right>$.}
\end{figure}

The difference between one particular element of the ensemble and the
average weighted by $W[\rho]$ is illustrated in figure \ref{fig:fluct}
for the quadratic form $\rho^2$.
Because the $\rho(\x_\perp)$ are uncorrelated at different points in the 
transverse plane of the nucleus, a particular configuration of $\rho$'s leads to a very
rough density profile; in contrast, the average smoothly follows the
Woods-Saxon density profile of a nucleus. This example simply
illustrates that the gradients are uncontrollably large for a given
configuration $\rho$ rendering any gradient expansion meaningless. On
the other hand, it is perfectly legitimate for ensemble averaged
quantities.

The typical momenta of ``hard'' particles is set by the saturation
scale which is of order $Q_s\sim 1$--$2$~GeV at RHIC energies; this scale
may be higher at the LHC. In contrast, the gradient
$\partial_{_X}$ for averaged quantities changes appreciably over
distance scales of the inverse nuclear radius given by $\sim
R_{_A}^{-1}\sim 40~$MeV for a large nucleus. The small magnitude of
this scale in the gradient expansion relative to the typical
saturation momentum justifies the gradient expansion for quantities
that are averaged over the ensemble of color charges.

\begin{figure}[htbp]
\begin{center}
\resizebox*{!}{5cm}{\rotatebox{90}{\includegraphics{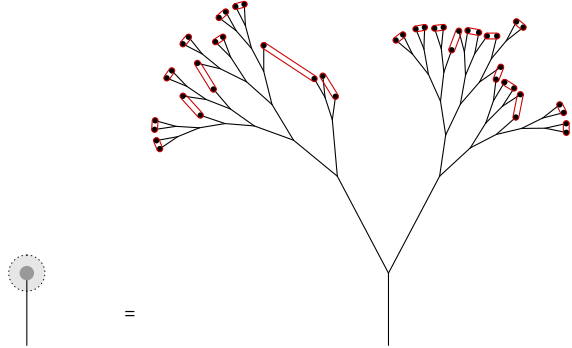}}}
\end{center}
\caption{\label{fig:tree_j}Example of a tree level contribution to the
average over the sources $j$ of the field expectation value for a
Gaussian distribution of sources. The links in red represent the
elementary correlators $\big<j(x)j(y)\big>$. The source connections represented here are for simplicity among nearest neighbors; all other
pairwise topologies are feasible.}
\end{figure}

The corresponding changes to the Feynman rules are described in
appendix \ref{sec:average}. Here it is sufficient to note that the
ensemble average is obtained by connecting all the external sources
$j$ in the manner specified by the distribution $W[\rho]$. For
instance, in the MV model $W[\rho]$ is a Gaussian, which implies that
all the sources must be connected pairwise. The objects
${\bs\Pi}^{_S}$, ${\bs\Pi}$ and ${\bs\Phi}$ that appear in the
Dyson-Schwinger equations (\ref{eq:C1}) and (\ref{eq:NC1}) must be
thought of as being averaged over $j$. In Feynman diagrams, we will
represent the average over $j$ by surrounding the diagram by a light
gray halo~: \setbox1\hbox to
0.8cm{\hfil\resizebox*{0.79cm}{!}{\includegraphics{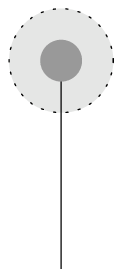}}}
\begin{equation}
\left<\vphantom{\Big[}\big<\phi(x)\big>\right>_j=\;\;\raise -7mm\box1\;\; .
\end{equation}
This compact notation encompasses a very large number of
contributions. For instance, at leading order, one would first
approximate $\big<\phi(x)\big>$ as the sum of all the tree diagrams,
an example of which is represented in figure \ref{fig:tree}. For each
such tree diagram, the sources $j$ (the black dots in figure
\ref{fig:tree}) are reconnected pairwise in all the possible ways. A
typical reconnection of the sources, corresponding to the topology of
figure \ref{fig:tree}, is displayed in figure \ref{fig:tree_j}.
Note that the ``loop order'' of a given diagram is a meaningful
concept only for diagrams {\sl before} they are averaged over
$j$. Indeed, as one can see by comparing the figures \ref{fig:tree}
and \ref{fig:tree_j}, the diagram before the $j$-average has 0 loops
and is of order $g^{-1}$. After the average is performed, while it has
a large number of ``loops" which do not contain any information about
the order in $g$ of the diagram.

\subsection{Boltzmann equation}
The final ingredient in the derivation of the Boltzmann equation is
the so-called ``quasi--particle ansatz'' which can be expressed as
\begin{eqnarray}
&&
{\bs G}_{-+}(X,p)=(1+f(X,\p)){\bs\rho}(X,p)\; ,
\nonumber\\
&&
{\bs G}_{+-}(X,p)=f(X,\p){\bs \rho}(X,p)\; ,
\end{eqnarray}
where the spectral function ${\bs\rho}(X,p)$ is
\begin{equation}
{\bs\rho}(X,p)\equiv {\bs G}_{_R}(X,p)-{\bs G}_{_A}(X,p)=
{\bs G}_{-+}(X,p)-{\bs G}_{+-}(X,p)\; .
\end{equation}
The physical assumption here is that the interactions in the system
are such that the collisional width of the dressed particles remains
small compared to their energy; the system is made up of long-lived
quasi--particles.

The Boltzmann equation can now be obtained as follows:
\begin{itemize}
  \item[{\bf i.}~] Write a Dyson-Schwinger equation analogous to
  eq.~(\ref{eq:CNC1}), but with the differential operator
  $\square+m^2+g{\bs\Phi}$ acting on the variable $y$ instead of $x$,
  and subtract it from eq.~(\ref{eq:CNC1}).
  \item[{\bf ii.}~] Rewrite this equation in terms of the Wigner
  transformed quantities and perform a gradient expansion 
  keeping only leading terms in $\partial_{_X}$.
  \item[{\bf iii.}~] Replace the Green's functions with the
  quasi-particle ansatz and drop the spectral function
  ${\bs\rho}(X,p)$ which appears as a factor in all the terms.
\end{itemize}
If the terms proportional to ${\bs \Pi}^{_S}$ were absent from
eq.~(\ref{eq:CNC1}), the steps outlined above would result in the
well-known Boltzmann--Vlasov equation,
\begin{eqnarray}
&&
2p\cdot \partial_{_X}f(X,\p)
+
g\partial_{_X}{\bs \Phi}(X)\cdot\partial_p f(X,\p)
=
\nonumber\\
&&\qquad\qquad=
(1+f(X,\p)){\bs \Pi}_{+-}(X,p)-f(X,\p){\bs \Pi}_{-+}(X,p)\; .
\label{eq:boltz0}
\end{eqnarray}
The extra term we have in the Dyson--Schwinger equations, proportional
to ${\bs \Pi}^{_S}$, will modify the Boltzmann--Vlasov equation. Two
key features of this novel term will prove essential in our
derivation. The first is that ${\bs\Pi}^{_S}(x,y)$ does not depend on
whether the points $x$ and $y$ are on the upper or lower branch of the
time contour. This is because the expectation value of the field, for
equal values of the sources $j_+$ and $j_-$, is the same on both
branches of the contour. The second feature is that the non-connected
part of the propagators drops out of the spectral function, for the
same reason. Hence, 
\begin{equation}
{\bs\rho}(X,p)={\bs G}_{-+}(X,p)-{\bs G}_{+-}(X,p)
={\bs G}_{-+}^{\rm c}(X,p)-{\bs G}_{+-}^{\rm c}(X,p)\; .
\end{equation}
Utilizing these two properties, we can perform the gradient expansion
for this additional term in the same way as performed for the usual
self-energy correction.  It modifies the right hand side of the
Boltzmann equation by an additive correction\footnote{Note that prior
to dropping the spectral function that appears in all terms, we would
have
\begin{eqnarray}
{\bs\Pi}^{_S}_{+-}(X,p){\bs G}_{-+}^{\rm c}(X,p)
-
{\bs\Pi}^{_S}_{-+}(X,p){\bs G}_{+-}^{\rm c}(X,p)
&=&
{\bs\Pi}^{_S}(X,p)
\Big[
{\bs G}_{-+}^{\rm c}(X,p)
-
{\bs G}_{+-}^{\rm c}(X,p)
\Big]
\nonumber\\
&=&
{\bs\Pi}^{_S}(X,p){\bs\rho}(X,p)\; .
\nonumber
\end{eqnarray}
} ${\bs\Pi}^{_S}(X,p)$. Therefore, our final expression for the
kinetic equation is
\begin{eqnarray}
&&
2p\cdot \partial_{_X}f(X,\p)
+
g\partial_{_X}{\bs \Phi}(X)\cdot\partial_p f(X,\p)
=
\nonumber\\
&&\qquad\qquad={\bs\Pi}^{_S}(X,p)
+
(1+f(X,\p)){\bs \Pi}_{+-}(X,p)-f(X,\p){\bs \Pi}_{-+}(X,p)\; .
\nonumber\\
&&
\label{eq:boltz}
\end{eqnarray}
The novel ``source term'' ${\bs\Pi}^{_S}(X,p)$ in this equation is
non-zero even if the particle distribution $f(X,\p)$ is zero. It is
therefore responsible for $f=0$ not being a fixed point of the above
equation;  the solution of this equation is
non-zero at later times even if the initial condition had a vanishing
particle distribution. In the next section, we will discuss further significant differences between this kinetic equation and the
conventional Boltzmann-Vlasov equation in eq.~(\ref{eq:boltz0}).

\section{Properties of the Glasma kinetic equation}
\label{sec:discussion}
In this section, we shall discuss the various terms in
eq.~(\ref{eq:boltz}) with emphasis on the differences between these
and those appearing in the conventional Boltzmann kinetic equation.
\subsection{Vlasov term}
We first consider the Vlasov term ($g\partial_{_X}{\bs\Phi}\cdot
\partial_p f$) in the Boltzmann equation.  We note that in performing
the average of the {\sl mean field} ${\bs \Phi}(X)$, over the external
sources $j$, the various correlation functions $\big<j(x_1)\cdots
j(x_n)\big>$ permitted by the distribution of sources $W[j]$ are nearly
translation invariant. The dependence of these correlators on the
barycentric co--ordinate $X\equiv (x_1+\cdots + x_n)/n$ is very slow
because it arises from the density profile of the colliding
nuclei\footnote{The fact that this density profile is not a constant
is the only effect in the problem that breaks translation
invariance.}. Therefore the 1-point function ${\bs\Phi}(X)$, averaged
over $j$, also has a very slow dependence on its argument $X$; its
Fourier transform with respect to $X$ has only modes with momenta on
the order of the inverse nuclear radius. As discussed previously, this
scale is very small relative to the typical momentum of the particles
under consideration and it is therefore legitimate to approximate it
as a Vlasov term.

As is well known, the effect of this term in the Boltzmann equation is
to change the momentum of particles as they move between regions where
the external field is different. Indeed, $g\partial_{_X}{\bs \Phi}$ is
the force that acts on the particles at point $X$ and accelerates them
towards regions of lower potential\footnote{For non central
collisions, the shape of the overlap region between the two nuclei is
elliptic; one has stronger gradients in the direction of the small
axis of the ellipsis relative to those in the direction of its large
axis. The Vlasov term therefore accelerates particles preferentially
in the direction of the small axis of the overlap region. This leads
eventually to elliptic flow and to an anisotropy of the spectrum of
particles in momentum space. This effect is obtained entirely within
kinetic theory without any assumption about the degree of
thermalization of the system.}. The mean field ${\bs\Phi}$ includes
not only the classical field directly produced by the external
sources, but also possibly a contribution coming from the particles
encoded in $f(X,\p)$. Such a modification may arise from a
modification of the particle dispersion relation due to the collective
action of the other particles. For instance, if the particles acquire
a medium mass with a weak space-time dependence, this mass can be
represented by a potential in the Vlasov term of the kinetic equation.

\subsection{Source term in the kinetic equation}
\label{sec:source_term}
Let us now consider the effect of the source term ${\bs\Pi}^{_S}(X,p)$
in eq.~(\ref{eq:boltz}), which can be obtained as the Wigner transform
of the product ${\bs S}(x){\bs S}(y)$. An interesting situation,
relevant for heavy ion collisions, is when tree diagrams are dominant
because the external source is strong ($gj\sim 1$). In this case, the
expectation value $\big<\phi(x)\big>$ is dominated by the retarded
classical field $\phi(x)$; the connected part of the 2-point function,
${\bs G}^{\rm c}$, is simply the propagator of a fluctuation on top of
the classical field,
\begin{equation}
\left({\bs G}^{\rm c}\right)^{-1} =\square+m^2+g\phi\; .
\end{equation}
One therefore immediately obtains the following
expression\footnote{This is the result for a potential $g
\phi^3/3!$. For an arbitrary potential $V(\phi)$, the expression of
${\bs S}(x)$ in this approximation would read
\begin{equation*}
{\bs S}(x)=j(x)-V^\prime(\phi(x))+\phi(x)V^{\prime\prime}(\phi(x))\; ,
\end{equation*}
where the prime denotes a derivative of the potential with respect to
$\phi$.} for ${\bs S}(x)$~:
\begin{eqnarray}
{\bs S}(x)&=&\Big[\square+m^2+g\phi(x)\Big]\,\phi(x)
\nonumber\\
&=&j(x)+\frac{g}{2}\phi^2(x)\; .
\label{eq:S-tree}
\end{eqnarray}
We see here that the effective source ${\bs S}(x)$ receives two
contributions~: 
\begin{itemize}
\item[{\bf i.}~] the external source $j(x)$ itself. This term is only
important if we want to use the Boltzmann equation in regions of space-time
where the external source is still active. In a heavy ion collision,
the color sources are present only on the light-cone at a proper time
$\tau=0$.  We will not consider this term further.
\item[{\bf ii.}~] A term quadratic in the classical field produced by
the external source; this term continues to contribute after the external
sources have stopped acting.
\end{itemize}
One may represent this effective source graphically as 
\setbox1\hbox to
4cm{\hfil\resizebox*{3.99cm}{!}{\includegraphics{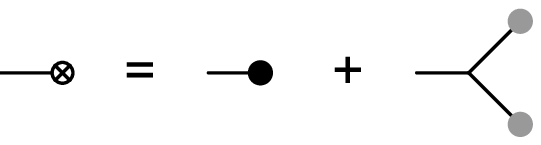}}}
\begin{equation}
{\bs S}(x)\equiv\;\;\; \raise -4.3mm\box1\;\;\; .
\end{equation}
The second term has a fairly straightforward interpretation. When the
term quadratic in $\phi$ in the classical equation of motion
\begin{equation}
\Big[\square+m^2\Big]\,\phi(x)
=j(x)-\frac{g}{2}\phi^2(x)\; ,
\end{equation}
is important, we see that the field is not a free field. If expanded
in particle modes, the number of particles in the field would change
with time. Therefore, if one switches between a description in terms
of classical fields to the kinetic equation at a stage where this
non-linear term is still significant, the source term in the Boltzmann
equation modifies the number of particles in order to take this effect
into account.

At tree level, the effective source ${\bs S}(x)$, and hence
${\bs\Pi}^{_S}$, is independent of the distribution of particles
$f$. As discussed previously, this is a straightforward consequence of
the fact that, at tree level, the 1-point function in the
Schwinger-Keldysh formalism can be rewritten entirely in terms of
retarded propagators that are $f$--independent. ${\bs\Pi}^{_S}$ is
therefore non-zero even if $f=0$. In contrast, the terms
${\bs\Pi}_{\pm\mp}$ in the r.h.s of the Boltzmann equation depend on
$f$ and vanish when $f=0$ as expected for {\sl collision terms}.
${\bs \Pi}^{_S}$ is therefore a {\sl source term} in
the Boltzmann equation, because it drives $f$ to a non-zero value even
if one has $f=0$ initially.

When we perform the average over $j$ of the disconnected product ${\bs
S}(x){\bs S}(y)$, we get both disconnected and connected source 
terms, \setbox1\hbox to
5.5cm{\hfil\resizebox*{5.49cm}{!}{\includegraphics{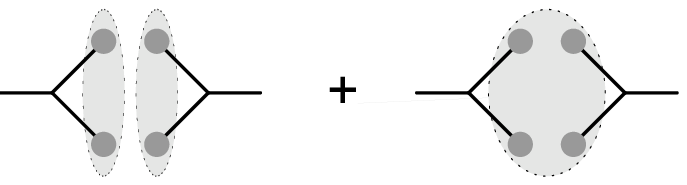}}}
\begin{equation}
\Big<{\bs S}(x){\bs S}(y)\Big>_j=\;\;\raise -7.5mm\box1\;\;.
\label{eq:Pis}
\end{equation}
depending on how the sources $j$ are reconnected. In this picture,
each light shaded area is simply connected after the average over $j$
has been performed, and all the sources $j$ it contains are linked in
all the possible ways that preserve its connectedness.

The first term in the r.h.s. of eq.~(\ref{eq:Pis}) corresponds to
contributions where we connect together only $j$'s that belong to
the same factor ${\bs S}$, $\big<{\bs
S}(x)\big>_j\big<{\bs S}(y)\big>_j$. Our previous remark about the
average over $j$ of the 1-point function ${\bs \Phi}(X)$ also applies
here to $\big<{\bs S}(x)\big>_j$~: its Fourier
transform only contains very soft modes of the order of the inverse of
the nuclear radius.  It is therefore nearly zero for the typical particle
momentum $p\sim Q_s$ we are interested in here. Thus only the
connected terms in the average of the source term $\big<{\bs S}(x){\bs
S}(y)\big>_j$ matter in the kinetic equation.

\subsection{Magnitude of field insertions}
\label{app:counting}
The source term in eq.~(\ref{eq:Pis}), as well as the other terms in
the right hand side of the Boltzmann equation, involve insertions of
the classical field $\phi(x)$. In this subsection, we present a simple
power counting that enables us to estimate the magnitude of such
insertions. To simplify the discussion, we shall assume that the
space--time coordinate $X$ corresponds to sufficiently late times when
the external source $j$ is zero and its influence is only felt through
the classical field $\phi(X)$ generated by the source at earlier
times.

Following the discussion after eq.~(\ref{eq:2p-corr}), we assume that there is hard momentum scale $Q_s$ in the
problem--the saturation scale in heavy ion collisions. Typical
particle momenta are of order $\p\sim Q_s$. In our toy model, the coupling constant $g$ 
has the dimension of a mass in 4 dimensions. To mimic the power counting in QCD, we will write it as
\begin{equation}
g\equiv\lambda Q_s ,
\end{equation}
where $\lambda$, like the QCD coupling constant, is dimensionless. We
assume that the coupling constant $\lambda \ll 1$.

To estimate the order of magnitude of the source term given
in eq.~(\ref{eq:Pis}), it is not sufficient to know the magnitude of
the classical field. Kinematical phase 
space constraints can alter the naive power counting. As these considerations will 
apply equally to the collision terms in the Boltzmann equation, it is worth our while to
discuss the power counting for the source term at length here.

From eq.~(\ref{eq:pi-source}) and eq.~(\ref{eq:Pis}), the naive power counting for the source term would 
give
\begin{equation}
{\bs\Pi}^{_S}(X,p)
=
\frac{\lambda^2 Q_s^2}{4}
\int d^4s \;e^{ip\cdot s}\;
\left<
\phi^2(X+\frac{s}{2})\phi^2(X-\frac{s}{2})
\right>_j\; . 
\label{eq:source-suppr}
\end{equation}
We will demonstrate that eq.~(\ref{eq:source-suppr})  vanishes when the momentum carried by the classical field $\phi$ is 
nearly on shell. Rewriting this expression entirely in momentum space in
terms of the Fourier transform ${\wt\phi}(k)$ of the classical field,
\begin{eqnarray}
&&
{\bs\Pi}^{_S}(X,p)
=
\frac{\lambda^2 Q_s^2}{4}
\int d^4s 
\int\frac{d^4k_1}{(2\pi)^4}\cdots\frac{d^4 k_4}{(2\pi)^4}\;
e^{ip\cdot s}
\nonumber\\
&&\!\!\!\!
\times
\;
e^{-ik_1\cdot(X+\frac{s}{2})}
e^{-ik_2\cdot(X+\frac{s}{2})}
e^{-ik_3\cdot(X-\frac{s}{2})}
e^{-ik_4\cdot(X-\frac{s}{2})}
\left<
{\wt\phi}(k_1){\wt\phi}(k_2){\wt\phi}(k_3){\wt\phi}(k_4)
\right>_j\; . 
\nonumber\\
&&
\label{eq:Pis1}
\end{eqnarray}
For the sake of simplicity, let us assume that the average over the
external source $j$ of the product of four fields factorizes into
products of averages of two fields as suggested by the source distribution in eq.~(\ref{eq:MV}). 
\begin{equation}
\left<
{\wt\phi}(k_1){\wt\phi}(k_2){\wt\phi}(k_3){\wt\phi}(k_4)
\right>_j
=
\left<
{\wt\phi}(k_1){\wt\phi}(k_3)\right>_j
\left<{\wt\phi}(k_2){\wt\phi}(k_4)
\right>_j+\mbox{other contractions}\; .
\end{equation}
For illustrative purposes, we consider only one of the possible
contractions corresponding to the connected topology of the second term in the r.h.s. of
eq.~(\ref{eq:Pis})). It is convenient at this point to denote
\begin{equation}
{\bs G}^{-+}_{\rm cl}(x,y)\equiv\left<\phi(x)\phi(y)\right>_j\; ,
\end{equation}
so that one has
\begin{equation}
\left<
{\wt\phi}(k_1){\wt\phi}(k_3)\right>_j
=
\int d^4Y\; e^{i(k_1+k_3)\cdot Y}\;
{\bs G}^{-+}_{\rm cl}\left(Y,\frac{k_1-k_3}{2}\right)\; .
\label{eq:Gclass}
\end{equation}
The definition of the object ${\bs G}^{-+}_{\rm cl}(x,y)$ is identical
to the usual definition of the $-+$ component of the Schwinger-Keldysh
propagators, except, as the notation suggests, it is constructed from the classical solution of
the equations of motion rather than from the full field operator. Inserting this definition into eq.~(\ref{eq:Pis1})
and keeping only the lowest order~\footnote{At
this order, this is equivalent to assuming, from the translational invariance in the transverse plane of 
a large nucleus,  that eq.~(\ref{eq:Gclass}) can be replaced by
\begin{equation*}
\left<
{\wt\phi}(k_1){\wt\phi}(k_3)\right>_j
\approx
(2\pi)^4\delta(k_1+k_3)
{\bs G}^{-+}_{\rm cl}(X,k_1)\; .\end{equation*}} in the gradients in $X$, one obtains
\begin{eqnarray}
{\bs\Pi}^{_S}(X,p)
&=&
\frac{\lambda^2Q_s^2}{4}
\int\frac{d^4k}{(2\pi)^4}\; 
{\bs G}^{-+}_{\rm cl}(X,k)\;{\bs G}^{-+}_{\rm cl}(X,p-k)
\nonumber\\
&&\qquad\qquad\qquad+\;\mbox{other contractions}\; .
\label{eq:Pis2}
\end{eqnarray}
Note that in this case there is only one other contraction, that leads
to the same contribution, thereby transforming the prefactor $1/4$
into a $1/2$. If the time $X^0$ at which this is evaluated is large
compared to $(Q_s)^{-1}$, the classical field that enters in the
definition of ${\bs G}^{-+}_{\rm cl}$ is mostly on-shell, and one can
write
\begin{equation}
{\bs G}^{-+}_{\rm cl}(X,k)
\approx 
2\pi\delta(k^2-m^2)\,f_{\rm cl}(X,\k)\; .
\label{eq:Gclass1}
\end{equation}
By analogy with eq.~\ref{eq:SK-f}, the distribution $f_{\rm cl}(X,\k)$ can be interpreted as representing
 the ``particle content'' of the classical field.  As eq.~(\ref{eq:Pis2}) has
exactly the structure of a $2\to 1$ collision term with on-shell
particles of equal mass, it is zero because of energy-momentum conservation. 

Therefore, to correctly estimate the magnitude of the source
term ${\bs\Pi^{_S}}$ when the classical field is weak, one
needs to properly account for the slight off-shellness of the field
Fourier modes. From the equation of motion
\begin{equation}
\frac{\square+m^2}{Q_s^2}\left(\frac{\phi}{\phi^*}\right)
+\frac{1}{2}\left(\frac{\phi}{\phi^*}\right)^2=0\; ,
\label{eq:EOM1}
\end{equation}
the off-shellness of the classical field comes from
its self-interactions. The simplest way to take this off-shellness into
account is to use the equation of motion in order to write
\begin{equation}
{\wt\phi}(k)=
\frac{\lambda Q_s}{2}\frac{1}{k^2-m^2}
\int\frac{d^4q}{(2\pi)^4}\;{\wt\phi}(q){\wt\phi}(k-q)\; ,
\end{equation}
and to replace some of the ${\wt\phi}$'s in eq.~(\ref{eq:Pis1}) by the
above relation. It is sufficient to replace two ${\wt\phi}$'s
in order to lift the kinematical constraints that came from the classical field having 
only nearly on-shell Fourier modes. This
substitution is straightforward. One obtains,
\begin{eqnarray}
&&{\bs\Pi}^{_S}(X,p)
=
\left(\frac{\lambda^2 Q_s^2}{4}\right)^2
\int\frac{d^4k}{(2\pi)^4}\frac{d^4q}{(2\pi)^4}
\frac{1}{(k^2-m^2)^2}
\nonumber\\
&&\quad\times\,
{\bs G}^{-+}_{\rm cl}(X,q)\;{\bs G}^{-+}_{\rm cl}(X,k-q)
\;{\bs G}^{-+}_{\rm cl}(X,p-k)
+\;\mbox{other contractions}\; .
\nonumber\\
&&
\label{eq:Pis3}
\end{eqnarray}

This contribution to the source term can be represented diagrammatically as 
\setbox1\hbox to 3.01cm{\hfil\resizebox*{3cm}{!}{\includegraphics{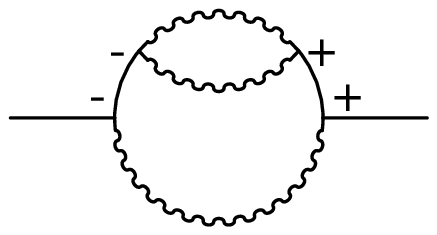}}}
\begin{equation}
{\bs\Pi}^{_S}(X,p)
=\;\;\;
\raise -7mm\box1\; ,
\end{equation}
where the solid lines represent ordinary vacuum propagators
($1/(k^2-m^2)$) and the wavy lines represent the correlation function
${\bs G}^{-+}_{\rm cl}$. It is interesting to note that this
contribution is identical in form to what one would have obtained in
the collision term of the conventional Boltzmann equation, except that
here the $G^{-+}$ propagators are made up of the classical fields.

We are now in a position to estimate the power counting of
contributions to the source term.  First, the order of magnitude of
the denominators $k^2-m^2$ is $Q_s^2$ because the momentum transfer
$k$ is of order $Q_s$ (and is not particularly close to the mass
shell). Each ${\bs G}^{-+}_{\rm cl}$ contains a delta function. Two of
them can be used to perform for free the integrations over the
energies $k^0$ and $q^0$, while the third provides the value of one
angular integration variable. We finally obtain the estimate
\begin{equation}
{\bs\Pi}^{_S}(X,p)\sim \frac{Q_s^2}{\lambda^2}
\left(\frac{n_{\rm cl}(X)}{n^*}\right)^2 
\frac{f_{\rm cl}(X,\overline{p})}{f^*}
\; ,
\label{eq:Pis-mag}
\end{equation}
where
\begin{equation}
n_{\rm cl}(X)\equiv \int\frac{d^3\k}{(2\pi)^3}\; f_{\rm cl}(X,\k)\; ,
\label{eq:nclass-def}
\end{equation}
$f^*\equiv \lambda^{-2}$, $n^*\equiv Q_s^3\lambda^{-2}$ and $n_{\rm
cl}(X)$ is the spatial density of particles corresponding to the
classical field. The expressions $f^*$ and $n^*$ correspond
respectively to the maximal values of $f_{\rm cl}$ and $n_{\rm cl}$
can have at early times $\lesssim Q_s^{-1}$). The argument
$\overline{p}$ cannot be specified exactly (in fact,
eq.~(\ref{eq:Pis-mag}) is an oversimplified version of the actual
formula for ${\bs\Pi}^{_S}$), but it is a momentum whose components are
of the same order of magnitude as those of $p$, the momentum of the
produced particle. This is an important point, because as time
increases, the support of $f_{\rm cl}$ shrinks in the $p_z$ direction
because of the longitudinal expansion of the system, thus making
$f_{\rm cl}(X,\overline{p})$ decrease as well (while in the center of
its support, it would stay constant).

Even if eq.~(\ref{eq:Pis3}) is not valid (say, if the average over $j$
were to generate connections among the fields that are not pairwise),
the estimate of ${\bs\Pi}^{_S}$ one obtains from it has a much wider
range of validity. (Eq.~(\ref{eq:Pis-mag}) is valid even in the
saturated regime.) We also note that as ${\bs\Pi}^{_S}$ is an
inhomogeneous term existing even when $f=0$, its magnitude depends
only on the time dependence of the classical field $\phi(x)$ through
$f_{\rm cl}$ and $n_{\rm cl}$.

\subsection{Collision terms}
The estimate of the various contributions to the collision term follow
very closely that of the source term. Let us start by listing the
terms we need to estimate. Because of the presence of the background
field and of the average over the external source $j$,
${\bs\Pi}_{-+}(X,p)$ can contain topologies that would not exist in
the vacuum. In fact, ${\bs\Pi}_{-+}(X,p)$ can contain terms that have
0, 1 and 2 loops\footnote{Naturally, there are also terms with an even
larger number of loops, but these are suppressed if the particle
occupation number is $f\ll \lambda^{-2}$.} before the average over the
external source is performed. We will denote by ${\cal C}_0[f], {\cal
C}_1[f]$ and ${\cal C}_2[f]$ their respective contributions to the
collision term. 

Let us start with ${\cal C}_0[f]$. Diagrammatically, it corresponds
to
\setbox1\hbox to 3.01cm{\hfil\resizebox*{3cm}{!}{\includegraphics{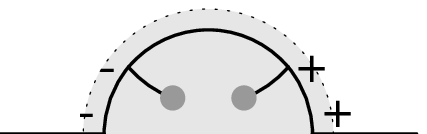}}}
\setbox2\hbox to 3.01cm{\hfil\resizebox*{3cm}{!}{\includegraphics{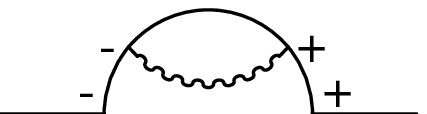}}}
\begin{equation}
{\cal C}_0[f]
=\;\;\;
\raise -7mm\box1=\;\;\;
\raise -7mm\box2\; .
\end{equation}
Note that this represents only one of the diagrams that
can possibly enter in ${\cal C}_0[f]$.  From the experience gained in
the estimate of the magnitude of ${\bs\Pi}^{_S}(X,p)$, we can readily
see that there must be at least four insertions of the classical field
for such a contribution to be kinematically viable when the classical
field becomes weak and has only near mass-shell Fourier modes. The
second equality shows one example of the topology one obtains after
the average over $j$. The corresponding expression reads
\begin{eqnarray}
&&{\cal C}_0[f]
=
\left(\frac{\lambda^2Q_s^2}{4}\right)^2
\int\frac{d^4k}{(2\pi)^4}\frac{d^4q}{(2\pi)^4}
\frac{1}{(k^2-m^2)^2}
\nonumber\\
&&\quad\times\,
{\bs G}^{-+}(X,q)\;{\bs G}^{-+}_{\rm cl}(X,k-q)
\;{\bs G}^{-+}_{\rm cl}(X,p-k)
+\;\mbox{other contractions}\; .
\nonumber\\
&&
\label{eq:C0}
\end{eqnarray}
The only difference between this expression and that of
${\bs\Pi}^{_S}$ in eq.~(\ref{eq:Pis3}) is that one of the correlators
${\bs G}^{-+}_{\rm cl}$ is now replaced by\footnote{This formula for
the correlator ${\bs G}^{-+}$ is only valid as long as the occupation
number $f(X,\p)$ is large compared to one. Its full expression contains
$\theta(p^0)+f(X,\p)$.}
\begin{equation}
{\bs G}^{-+}(X,p)=2\pi\delta(p^2-m^2)f(X,\p)\; ,
\end{equation}
that involves the distribution $f(X,p)$ rather than the classical
distribution $f_{\rm cl}(X,p)$. From this analogy, we can estimate the
magnitude of ${\cal C}_0[f]$ directly from that of ${\bs\Pi}^{_S}$ in
eq.~(\ref{eq:Pis-mag}), by substituting one factor $f_{\rm cl}$ or
$n_{\rm cl}$ by respectively $f$ or $n$. Here $n$ is the spatial
density defined from $f$ in the same way as in
eq.~(\ref{eq:nclass-def}). We obtain 
\begin{equation}
{\cal C}_0[f]\sim \frac{Q_s^2}{\lambda^2}
\Big[
\left(\frac{n_{\rm cl}(X)}{n^*}\right)^2 \frac{f(X,\overline{p})}{f^*}
\oplus
\frac{n_{\rm cl}(X)}{n^*}
\frac{n(X)}{n^*}
\frac{f_{\rm cl}(X,\overline{p})}{f^*}
\Big]
\; .
\label{eq:C0-mag}
\end{equation}

Similarly, ${\cal C}_1[f]$ corresponds to diagrams of the type 
\setbox1\hbox to 3.01cm{\hfil\resizebox*{3cm}{!}{\includegraphics{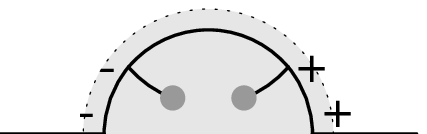}}}
\setbox2\hbox to 3.01cm{\hfil\resizebox*{3cm}{!}{\includegraphics{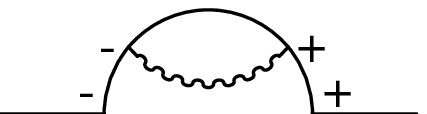}}}
\begin{equation}
{\cal C}_1[f]
=\;\;\;
\raise -7mm\box1=\;\;\;
\raise -7mm\box2\; ,
\end{equation}
and the corresponding expression reads
\begin{eqnarray}
&&{\cal C}_1[f]
=
\left(\frac{\lambda^2Q_s^2}{4}\right)^2
\int\frac{d^4k}{(2\pi)^4}\frac{d^4q}{(2\pi)^4}
\frac{1}{(k^2-m^2)^2}
\nonumber\\
&&\quad\times\,
{\bs G}^{-+}(X,q)\;{\bs G}^{-+}_{\rm cl}(X,k-q)
\;{\bs G}^{-+}(X,p-k)
+\;\mbox{other contractions}\; .
\nonumber\\
&&
\label{eq:C11}
\end{eqnarray}
Here we replace two out of three correlators ${\bs G}^{-+}_{\rm cl}$
by ${\bs G}^{-+}$; the power counting for this diagram is then 
\begin{equation}
{\cal C}_1[f]\sim \frac{Q_s^2}{\lambda^2}
\Big[
\frac{n_{\rm cl}(X)}{n^*}\frac{n(X)}{n^*} \frac{f(X,\overline{p})}{f^*}
\oplus
\left(\frac{n(X)}{n^*}\right)^2
\frac{f_{\rm cl}(X,\overline{p})}{f^*}
\Big]
\; .
\label{eq:C1-mag}
\end{equation}

Finally, for the 2-loop contribution to the collision term, we
have
\setbox1\hbox to 3.01cm{\hfil\resizebox*{3cm}{!}{\includegraphics{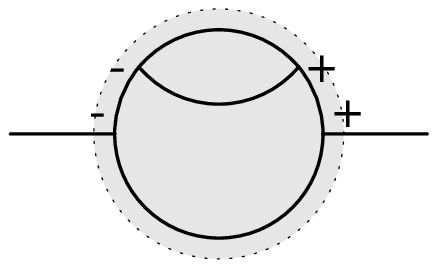}}}
\setbox2\hbox to 3.01cm{\hfil\resizebox*{3cm}{!}{\includegraphics{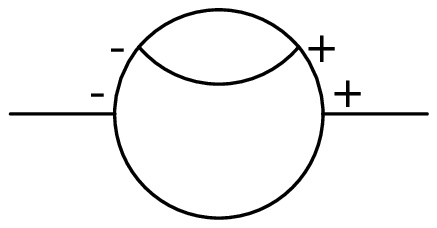}}}
\begin{equation}
{\cal C}_2[f]
=\;\;\;
\raise -7mm\box1=\;\;\;
\raise -7mm\box2\; ,
\end{equation}
\begin{eqnarray}
&&{\cal C}_2[f]
=
\left(\frac{\lambda^2Q_s^2}{4}\right)^2
\int\frac{d^4k}{(2\pi)^4}\frac{d^4q}{(2\pi)^4}
\frac{1}{(k^2-m^2)^2}
\nonumber\\
&&\quad\times\,
{\bs G}^{-+}(X,q)\;{\bs G}^{-+}(X,k-q)
\;{\bs G}^{-+}(X,p-k)
+\;\mbox{other contractions}\; ,
\nonumber\\
&&
\label{eq:C2}
\end{eqnarray}
and
\begin{equation}
{\cal C}_2[f]\sim \frac{Q_s^2}{\lambda^2}
\left(\frac{n(X)}{n^*}\right)^2
\frac{f(X,\overline{p})}{f^*}
\; .
\label{eq:C2-mag}
\end{equation}

\subsection{Discussion}
Following the power counting in equations (\ref{eq:Pis-mag}),
(\ref{eq:C0-mag}), (\ref{eq:C1-mag}) and (\ref{eq:C2-mag}), we are now
in a position to discuss qualitatively the relative magnitude of the
various terms at different stages of the evolution of the system. An
important facet of the temporal evolution is that the functions
$f_{\rm cl}$ and $n_{\rm cl}$ are determined once and for all from the
classical field $\phi(x)$ itself. They do not receive any feedback
from the particle distributions or densities, denoted by $f$ and $n$
respectively, that are created in the evolution by the source term
${\bs\Pi}^{_S}$. The time dependence of $n_{\rm cl}(X)$ is driven by
the expansion of the system; therefore at times larger than
$(Q_s)^{-1}$, one has
\begin{equation}
\frac{n_{\rm cl}(X)}{n^*}\sim \frac{1}{Q_s\tau}\; .
\end{equation}
This reduction of the classical particle density $n_{\rm cl}$ with
time happens because the support in momentum space of the
corresponding phase-space density $f_{\rm cl}$ shrinks. At a given
space-time location $X$ (specified by the space-time rapidity $\eta$),
only particles with a matching momentum rapidity $y=\eta$ can stay
for a long time. Therefore, inside its support, $f_{\rm cl}$ remains
constant satisfying 
\begin{equation}
\frac{f_{\rm cl}(y\approx\eta)}{f^*}\sim 1\; .
\end{equation}
Note that at times smaller than $(Q_s)^{-1}$, ${f_{\rm cl}}/{f^*}$ and
${n_{\rm cl}(X)}/{n^*}$ are also both of order 1 because the classical
field is completely saturated.

However, in all the estimates of the previous subsection, $f_{\rm cl}$
is evaluated at some arbitrary location $X$ and momentum
$\overline{p}$. Therefore, $\overline{p}$ will eventually fall outside
of the support of $f_{\rm cl}$, and $f_{\rm cl}$ will decrease
quickly\footnote{The precise time dependence of this fall depends on
the $p_z$ dependence of $f_{\rm cl}$. To take an extreme case, there
would be no fall at all if $f_{\rm cl}$ is independent of $p_z$.}
after that happens. For $f_{\rm cl}$, which comes entirely from the
classical field $\phi$, the only time-scale in the problem is $1/Q_s$
and thus we expect $f_{\rm cl}$ to start decreasing at times larger
than $1/Q_s$.

At early times, $\tau\to 0$, the system does not have particles yet
and we have $f=n=0$. Obviously, in this regime, only the source term
${\bs\Pi}^{_S}$ is important in the right hand side of the Boltzmann
equation. The corresponding physics is that a population of particles,
described by the occupation number $f$, is built up from the decay of
the classical field. However, these particles are still too few to
have collisions at a significant rate. Eq.~(\ref{eq:Pis-mag}) tells us
that ${\bs\Pi}^{_S}\sim \frac{Q_s^2}{\lambda^2}$ in this regime.

As a rough estimate, if we integrate this source term in the range
$0\le\tau\le Q_s^{-1}$, we find that the occupation number for
particles of momentum $p\sim Q_s$ at a time $\tau\sim Q_s^{-1}$ is
\begin{equation}
\frac{f(\tau=Q_s^{-1})}{f^*}\sim 1\; .
\end{equation}
At this time, all the components of the momenta of these particles are
typically of order $Q_s$. Therefore, we also have
\begin{equation}
\frac{n(\tau=Q_s^{-1})}{n^*}\sim 1\; .
\end{equation}
At times around $(Q_s)^{-1}$ all the terms in the right side of the
Boltzmann equation are of equal magnitude. Indeed, in this regime,
terms with an arbitrarily large number of loops contribute equally to
the collision term when $f\sim f^*$.  There would therefore be an
equally large ${\cal C}_3[f], {\cal C}_4[f], $~etc... In practice,
this means that one should start using the Boltzmann equation only at
later times.

At later times, $\tau\ge Q_s^{-1}$, collisions among the particles
become important and their qualitative effect is to broaden the
momentum distribution of the particles represented by $f$, thereby
counteracting the effect of the expansion\footnote{In the absence of
collisions, $f$ would be affected by the system expansion in a similar
way to $f_{\rm cl}$, and its support would shrink like $\tau^{-1}$ in
the $p_z$ direction.}  of the system. Thanks to these collisions,
$f(X,\overline{p})$ falls at a lesser rate compared to $f_{\rm
cl}(X,\overline{p})$ (which is not affected by collisions), which
eventually leads to the dominance of ${\cal C}_2[f]$ over all the
other terms in the right hand side of the Boltzmann equation. When
this occurs, our Boltzmann equation is identical to the usual one.
The detailed mechanisms of this transition between the classical field
dominated regime and the kinetic regime will be discussed in a future
work. In particular, it will be interesting to compare, for the QCD
case, the temporal evolution of the kinetic equation for the glasma
with the ``bottom up" scenario of thermalization~\cite{BaierMSS1}.

\section{Summary and Outlook}

In this work, we developed the formalism of
Refs.~\cite{GelisV2,GelisV3} for particle production in the presence
of strong sources to construct a kinetic theory relevant for the early
``glasma" stage of a heavy ion collision. In particular, we considered
for simplicity, the dynamics of a $\phi^3$ theory in the presence of
strong sources. Much of our discussion however is completely general
and could in principle be extended to describe the dynamics of gauge
fields exploding into the vacuum after a heavy ion collision. We
showed that the relevant kinetic equation for the particle
distributions $f$ has the structure of a Boltzmann equation with an
additional inhomogeneous ($f$-independent) source term denoting
particle creation from the decay of the classical field. The collision
terms in the Boltzmann equation also have novel features. In addition
to the usual contribution from the two loop self energy, there are
0-loop and 1-loop contributions that affect the particle phase space
distributions. We outlined the power counting that controls the
magnitude of the contributions of the source term and the collision
terms. The temporal evolution of these contributions was discussed
only briefly and will be discussed in detail elsewhere.

There are several unresolved issues that should be addressed in future
work. Primarily, we would like to understand precisely how the
derivation here plays out in the QCD case. In
Refs.~\cite{ArnolLM1,ArnolLMY1,RebhaRS1,RebhaRS2}, it was shown that
instabilities of the Weibel
type~\cite{Mrowc1,Mrowc2,Mrowc3,Mrowc4,Mrowc5,RandrM1} can spoil the
bottom up scenario of thermalization. Such an instability is also seen
in the CGC framework in the explosive growth of small fluctuations
about the classical background field~\cite{RomatV1,RomatV2,RomatV3}
and has a natural interpretation as quantum fluctuations about the
classical background fields on the light cone~\cite{FukusGM1}. A
numerical study of instabilities in a field+particle framework has
been performed~\cite{DumitNS1} but we would like to better understand
how the effects of such instabilities manifest themselves in the
kinetic equation for the glasma. It would be especially interesting to
uncover whether Kolmogorov turbulent spectra~\cite{ZakhaLF1} arise as
a consequence of these instabilities~\cite{ArnolM1,MuellSW2} and
whether this phenomenon of ``turbulent thermalization" can be
accommodated in our kinetic framework.

\section*{Acknowledgements}
We would like to thank K. Fukushima, T. Lappi, L. McLerran and
A. H. Mueller for enlightening discussions.  FG and RV would like to
thank McGill University and the Galileo Galilei institute in Florence
and INFN for their kind hospitality in the course of this work. RV's
research was supported by DOE Contract No.  DE-AC02-98CH10886. S.J. is
supported in part by the Natural Sciences and Engineering Research
Council of Canada.

 \appendix

\section{Average over the sources}
\label{sec:average}
We have seen that it is crucial for the validity of the gradient
expansion to consider quantities averaged over the source $j$ coupled
to the fields. We shall discuss briefly here how this average can be
accounted for in our formalism.

Let us start from the generating functional for Green's functions of
the Schwin\-ger-Keldysh formalism\footnote{In order to keep the
notations compact, we denote by a boldface letter ${\bs\eta}$ the pair
${\bs\eta}\equiv(\eta_+,\eta_-)$, where the $\pm$ indices refer to the
Schwinger-Keldysh closed time path.} $Z_j[{\bs\eta}]$, {\sl for a
given configuration $j$ of the external source}. We define it in such
a way that the $n$-point Green's functions is obtained by
differentiating $n$ times with respect to ${\bs\eta}$, and then by
setting the auxiliary source ${\bs\eta}$ to zero. From what we have
said in section \ref{sec:nbar}, this generating functional is related
to the sum of all the vacuum-vacuum diagrams by~:
\begin{equation}
Z_j[{\bs\eta}]=e^{i{\cal V}_{_{SK}}[j+{\bs\eta}]}\; ,
\end{equation}
where we have again used a compact notation compared to
eq.~(\ref{eq:G-full}). We do not use a boldface letter for the
external source $j$, in order to emphasize the fact that it is
identical on both branches of the closed time path.

From this object, it is very easy to construct the generating
functional for Green's functions that are averaged over some ensemble
of external sources, with a distribution $W[j]$, as~:
\begin{equation}
Z[{\bs\eta}]=\int \big[Dj\big]\;W[j]\;e^{i{\cal V}_{_{SK}}[j+{\bs\eta}]}\;.
\label{eq:Z}
\end{equation}
In order to see how this average over $j$ can be accounted for in the
Feynman rules, it is useful to write the generating functional for a
fixed $j$ as follows~:
\begin{eqnarray}
&&
e^{i{\cal V}_{_{SK}}[j+{\bs\eta}]}
=
\exp
\Big(i\int_{\cal C}d^4x\, V\left(\frac{\delta}{\delta{\bs\eta}(x)}\right)\Big)
\nonumber\\
&&\quad\qquad\times
\exp\Big({-\frac{1}{2}\int_{\cal C}d^4x d^4y\,
(j(x)+{\bs\eta}(x)){\bs G}^0(x,y)(j(y)+{\bs\eta}(y))
}\Big)\; ,
\label{eq:Zj}
\end{eqnarray}
where $V$ is the sum of all the interaction terms in the theory under
consideration (i.e. all the terms of the Lagrangian density that are
of degree $\ge 3$ in the field). In this formula, ${\bs G}^0(x,y)$
denotes the free propagator in the Schwinger-Keldysh formalism (as
opposed to the full propagator defined in eq.~(\ref{eq:G-full})).  It
is now convenient to write the second exponential in the r.h.s. of
eq.~(\ref{eq:Zj}) as the action of a translation operator on a
functional that does not depend on $j$,
\begin{eqnarray}
&&
\exp\Big({-\frac{1}{2}\int_{\cal C}d^4x d^4y\;
(j(x)+{\bs\eta}(x)){\bs G}^0(x,y)(j(y)+{\bs\eta}(y))
}\Big)
=
\nonumber\\
&&\quad=
\exp\Big({i\int_{\cal C}d^4z\, j(z)\frac{\delta}{\delta{\bs\eta}(z)}}\Big)
\;
\exp\Big({-\frac{1}{2}\int_{\cal C}d^4x d^4y\;
{\bs\eta}(x){\bs G}^0(x,y){\bs\eta}(y)
}\Big)\; .
\nonumber\\
&&
\end{eqnarray}
By inserting this formula in eq.~(\ref{eq:Zj}), and then in
eq.~(\ref{eq:Z}), we obtain the following expression~:
\begin{eqnarray}
&&
Z[{\bs\eta}]
=
\left\{
\int \big[Dj\big]\;W[j]\;
e^{i\int_{\cal C}d^4z\, j(z)\frac{\delta}{\delta{\bs\eta}(z)}}
\right\}
\nonumber\\
&&\qquad\times
\exp
\Big(i\int_{\cal C}d^4x\, V\left(\frac{\delta}{\delta{\bs\eta}(x)}\right)\Big)
\;
\exp\Big({-\frac{1}{2}\int_{\cal C}d^4x d^4y\;
{\bs\eta}(x){\bs G}^0(x,y){\bs\eta}(y)
}\Big)\; .
\nonumber\\
&&
\label{eq:Z1}
\end{eqnarray}
The terms on the second line are nothing but the generating functional
for the same theory {\sl without any external source} (since it does not
depend on $j$). As we can see, the effect of the
average over the external source $j$ is to bring a prefactor which is
a certain functional of the operator $\delta/\delta{\bs\eta}$. Such a
term can be interpreted as additional couplings among the fields,
since one can always write~:
\begin{equation}
\left\{
\int \big[Dj\big]\;W[j]\;
e^{i\int_{\cal C}d^4z\, j(z)\frac{\delta}{\delta{\bs\eta}(z)}}
\right\}
\equiv 
\exp\Big({i\int_{\cal C}d^4x\;
U\left(\frac{\delta}{\delta{\bs\eta}(x)}\right)}
\Big)\; .
\label{eq:U}
\end{equation}
 What this derivation makes obvious is that, for calculating averaged
quantities over the ensemble of external sources $j$, one can forget the
external sources altogether, and include additional vertices to the
theory\footnote{Note that formally, the ``potential'' $U$ is the
connected part of the Fourier transform of the functional $W[j]$. This
means that in the particular case where $W[j]$ is a Gaussian, there is
only one coupling in the new potential $U$, which couples two
fields. Because ``interaction terms'' that are quadratic in the fields
can in general be handled in closed form, it is possible in this case
to absorb the potential $U$ in a modification of the propagator ${\bs
G}^0(x,y)$ in eq.~(\ref{eq:Z1}).}, as prescribed by
eq.~(\ref{eq:U}). Note that this is equivalent to calculating a
quantity in an arbitrary $j$, and then reconnecting all the $j$'s
among themselves in all the possible ways permitted by $\ln(W[j])$.

\bibliographystyle{unsrt}

\end{document}